# PREDICTING CLIENT SATISFACTION THROUGH (E-MAIL) NETWORK ANALYSIS: THE COMMUNICATION SCORE CARD

Dirk Brunnberg
Zeppelin University
Friedrichshafen,
Germany
d.brunnberg@zeppelin-university.net

Peter A. Gloor
MIT CCI
5 Cambridge Center
Cambridge MA 02138
pgloor@mit.edu

Gianni Giacomelli
genpact
New York, USA
gianni.giacomelli@genpact.com

**ABSTRACT**

This study seeks to better understand the network characteristics of client support teams by analyzing the teams' e-mail communication networks and comparing it to client organization's satisfaction. In collaboration with a large service provider we studied the impact of network properties on the satisfaction of client organizations. In particular, we found that social network metrics correlate with client satisfaction as measured by Net Promoter Score (NPS). A Communication Score Card is suggested as a dashboard to continuously measure client satisfaction, illustrating that data-driven analysis might help improving service providers' service quality management.

**INTRODUCTION**

Social Network Analysis as a tool for organizational theory and research dates back to the 1930s (Jack 2010). It allows us to understand social mechanisms and their impact on outcomes in any kind of organization. Granovetter (2005) posits that Social Network Analysis is able to reveal the key drivers of economic and organizational action: *information*, *sources of reward and punishment* as well as *trust* between people. Following in this tradition, our study tries to extract actionable information from e-mail communication in organizations.

These insights are of high interest in different industries. The organization we studied is a large service provider active in an information-based industry, where communication is preferably exchanged via e-mail. Face-to-face communication and meetings with the client employees are not possible on a daily basis, making e-mail a key communication channel.

The service provider gives B2B-Support, such as Financial & Account (F&A) support and other types of back and middle office support. Normally, client organization's satisfaction and the quality and effectiveness of the services projects are measured through surveys, but these are neither timely nor precise enough to provide detailed information about specific projects and tasks. We propose to use Social Network Analysis (SNA) for tracking the communication patterns between client and provider work team members.

There is a broad stream of research that tries to find correlations between social network structure and performance (Gloor 2005). While initially most social network analysis has been conducted by surveys filled out manually by participants (Cummings & Cross 2003), recently studies using e-mail networks (Aral & Van Alstyne 2007) have become popular.

Most of these studies attest a positive relationship between *network metrics*, such as betweenness or degree centrality and the performance of individuals, work teams and organizations (Bulkley & Van Alstyne, 2006; Gloor, Paasivaara, Schoder, & Willems, 2008). In earlier research it was found that teams are more creative the higher their (pre-existing) social capital is (Gloor et al., 2012; Nemoto, Gloor, & Laubacher, 2011). In other work, it was shown that structural properties of social networks are associated with performance. For example, work teams have lower outcomes in more core-periphery and hierarchical structure as well as when leaders have 'structural holes' within their networks[1] (Cummings & Cross, 2003).

This study intends to extend this stream of research by (1) deriving empirical insights from prior theory and research and (2) developing a 'Network Communication Score Card' that consists of several metrics and allows to predict clients' satisfaction.

**DATA GENERATION & PREPARATION**

In order to find implications for measuring client work teams' satisfaction through (e-mail) network analysis, the e-mail communication of 38 work teams was collected through the organization's firewall log by including all e-mails sent by the provider's account executives, received by them, or where they were cc'd for the 38 accounts. The organization's clients are mostly multinational corporations. The time period of e-mail collection of the communication between provider work teams and clients' employees was from June until December 2012. Because of data quality issues, in the end we were only able to use 13 out of the 38 data sets for our analysis, because on the one hand there were large holes in the e-mail data we collected. For instance, there were months without e-mail for some of the accounts. On the other hand, we also had quality problem with our dependent variable NPS, see below in the section "Measuring Client Satisfaction".

---

[1] For further explanation of 'structural holes' and their implication for leadership and organizations, see Burt (2001).

## CONCEPTUAL FRAMEWORK

We propose a framework based on metrics in five dimensions, which are related to network (1) structure, (2) flow, (3) dynamics and (4) evolution (Bulkley & Van Alstyne, 2006). In addition, we conducted a (5) sentiment analysis of the e-mails' subject lines that indicates whether emotionality of language is positive or negative for client organization's satisfaction.

Subsequently the dependent variable, which measures client satisfaction, will be explained. After that, the Network Communication Score Card's relevant network metrics will be introduced based on prior research.

### Measuring Client Satisfaction

The organization has been measuring service quality and clients satisfaction for many years using the Net Promoter Score (NPS) to examine their projects' success and client approval. NPS asks individuals at the client organization a straightforward question: "how likely are you to recommend our services to a friend?" The NPS score is a simple metric calculated by assigning responses from 0-10 that group client employees in three classes: promoters (9-10: extremely satisfied client employee), passively satisfied (7-8 score) and detractors (0-6 ratings: not likely to give recommendation). The NPS score is the result of subtracting the percentage of detractors from the percentage of promoters (Reichheld, 2003). For this study, the NPS responses for the fourth quarter of 2012 were used.

Additionally, an internal set of eight quality metrics was used by aggregating questions which were also asked to the same client employees. For example, satisfaction about issues such as the management of people or process communication was gathered. Then, for every work team the average value of these eight values, obtained in quarter four of 2012, was calculated and named as Key Performance Drivers (KPD). Out of the 38 accounts, however, we were only able to use 14, because for some accounts only a single client employee had answered the NPS and KPD questions, leading to a non-representative sample. We therefore only included the accounts in our analysis where more than 20 individuals from the clients' workforce had answered the questions.

## NETWORK METRICS & HYPOTHESES

As earlier research has shown, several assumptions can be made related to network structure and organization's performance. The network metrics presented in the following are explained by Wasserman & Faust (1994). The correlation hypotheses are:

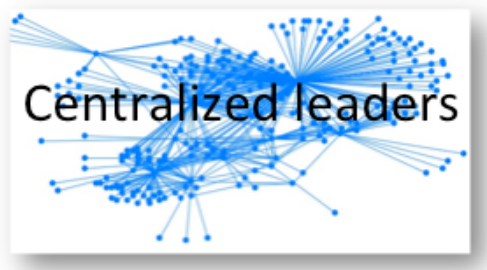 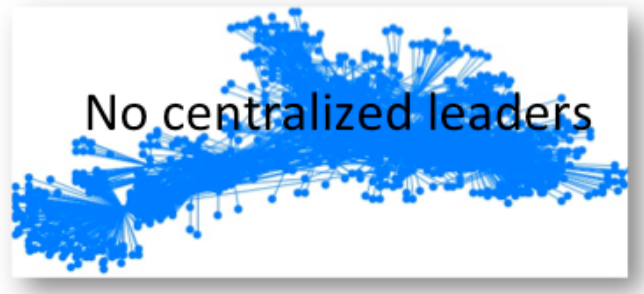

**Figure 1: Centralized (left) and non-centralized (right) networks**

### (1) Network Centrality, Density & Continual Network Structure

Based on prior research, network centrality, i.e. a few individuals significantly more central than the rest of the actors in the network, seems to be an indicator for successful individuals, teams and organizations (Bulkley & Van Alstyne, 2006; Gloor et al., 2008). Taking betweenness and degree centrality as our relevant social network metrics, we speculate:

*H1: The higher centrality of the work teams' network, measured by betweenness and degree centrality, the higher is client satisfaction.*

In addition, prior research showed that higher network density leads to higher performance in an open source development community (Kidane & Gloor, 2007). This is consistent with the findings of Cummings & Cross (2003) and leads to:

*H2: The denser the communication network, the higher is client organization's satisfaction.*

As structural cohesion, flat hierarchies and well-organized information flow seem to have an impact on individual and organizational performance (Cummings & Cross, 2003), it was examined how many new actors sending and receiving e-mail appear every month from June until December 2012 as well as on average. As people dislike change, we hypothesize:

*H3: The addition of new actors to the network has a negative impact on client organization's satisfaction.*

**(2) Leadership Oscillation**

Earlier research showed that rotating/oscillating leadership within teams can be a predictor for creativity of open source programmers (Kidane & Gloor, 2007). Figure 2 illustrates the differences between oscillating leadership networks and networks with steady leadership. For clients' employees, one can expect that having the same leader will convey a feeling of consistency and clear role allocation. Prior research additionally showed that there are divergent network mechanisms depending on whether the task is performance- or creativity-oriented. Therefore, we hypothesize that rotating leadership lowers client employees' satisfaction:

*H4: Higher leadership oscillation in work teams leads to less client organization's satisfaction.*

**(3) Responsiveness**

Regarding the communication behavior of work team members several metrics were gathered. On the one hand, the Average Response Time (ART) as well as the median response time was calculated for every work team over the entire time period. It was found in prior research that the faster people respond to e-mail on average, the happier they are (Merten & Gloor, 2010). Fast e-mail reply is also an indicator for high-quality work and fast information exchange (Aral & Van Alstyne, 2007). Another indication was shown in frequency of communication: the more frequent (short) messages are exchanges, the higher was the output of individuals (Bulkley & Van Alstyne, 2006). Therefore, we posit:

*H5: The faster e-mails are answered, the higher is the client organization's satisfaction.*

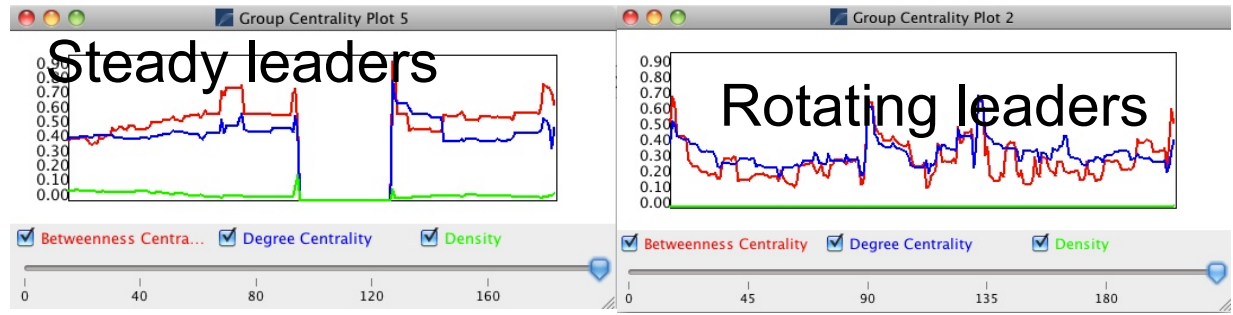

Figure 2: Steady leaders (left) and oscillation in leadership (right) over time (x-axis) plotted against changing betweenness centrality (y-axis).

**(4) Variation in Contribution Index**

The Average Weighted Variance in Contribution Index (AWVCI) has been repeatedly used in prior research projects. It is an indicator for a team's communication balance. The calculation of the contribution index is as follows:

$$contribution\ index = \frac{messages_{sent} - messages_{received}}{messages_{sent} + messages_{received}}$$

The contribution index ranges from -1 to +1, where it is +1 if a person only sends and -1 if a person only receives e-mails (Gloor et al. 2003). The AWVCI accounts for high variances in the contribution index by weighting it "with the number of total edges on that particular day" (Gloor et al., 2008).

Previous research indicates that lower variation in contribution index results in higher team creativity (P. Gloor et al., 2008). As we are looking for performance, and not creativity, it is expected that the higher AWVCI will lead to more client satisfaction.

*H6: The higher values of AWVCI are, the more satisfied are client's employees.*

**(5) Emotionality & "Honest Signals"**

In face-to-face communication, the emotionality and sentiment of messages can be interpreted by the counterpart's body language. In e-mail communication, it is harder to assess these "honest signals" (Pentland 2008). This study extends results from prior research and conducts an automatic sentiment analysis of every work team's e-mail subject line (Gloor et al. 2012). This way, it is possible to measure emotional and non-emotional language, acting as "honest signals" of e-mail communication. As client employees value clear and factual language over overly positive "slang", it is assumed that "honest and transparent language" will lead to higher satisfaction measures, therefore:
*H7: The less emotional work teams' e-mail communication is, the higher is client employees' satisfaction.*

## RESULTS

For most hypotheses, we find significant correlations, demonstrating that indeed network structure and dynamics are able to predict client organization's satisfaction.

### Centrality & Structure

Hypotheses 1, 2 & 3 assumed that centrality as well as density has positive effects on client employees' satisfaction. We find indeed significant Pearson correlation for the KPD Q4 variable.
This means that the more a work team is centrally led, the higher is the clients' satisfaction. Correspondingly, significant negative correlations are found with increasing numbers of new team members. Therefore, the more central a work team is being lead, and the more stable and consistent a work team operates, the higher is client organization's satisfaction. Hypotheses 1, 2 & 3 are therefore confirmed.

### Leadership Oscillation & Contribution Variance

The theory on leadership and networks leads us to expect different results on oscillation being positive or negative depending on the task. This study showed a significant negative correlation ($p<.05$) with client organization's satisfaction (see Figure 3). This is consistent with hypotheses 1, 2 & 3 as it underlines the importance of clear communication with the clients' employees from always the same and steady account managers and leaders.
Correlations between AWVCI and client satisfaction and KPD are (almost) significant ($p<0.085$).

| Hypotheses | H 1, 2 & 3 | | | | H4 | H5 | H6 | H7 |
|---|---|---|---|---|---|---|---|---|
| | Avg GBC | Avg GDC | Avg Density | Avg. New Actors | Sum of Oscillation | ART Median | AWVCI (weighted by #actors) | Emotionality (cumulated pos. sentiment) |
| **NPS Q4** | | | | | | | | |
| Pearson | .503 | .454 | .402 | -.454 | -.604* | -.414 | .418 | -.44 |
| Sig. (2-tailed) | 0.08 | 0.119 | 0.173 | 0.119 | 0.029 | 0.159 | 0.156 | 0.132 |
| N | 13 | 13 | 13 | 13 | 13 | 13 | 13 | 13 |
| **KPD Q4** | | | | | | | | |
| Pearson | .645* | .609* | .496 | -.579* | -.644* | -.533 | .495 | -.572* |
| Sig. (2-tailed) | 0.017 | 0.027 | 0.085 | 0.038 | 0.018 | 0.061 | 0.085 | 0.041 |
| N | 13 | 13 | 13 | 13 | 13 | 13 | 13 | 13 |

*. Correlation is significant at the 0.05 level (2-tailed)

**Table 1: Statistics of Correlation Hypotheses**

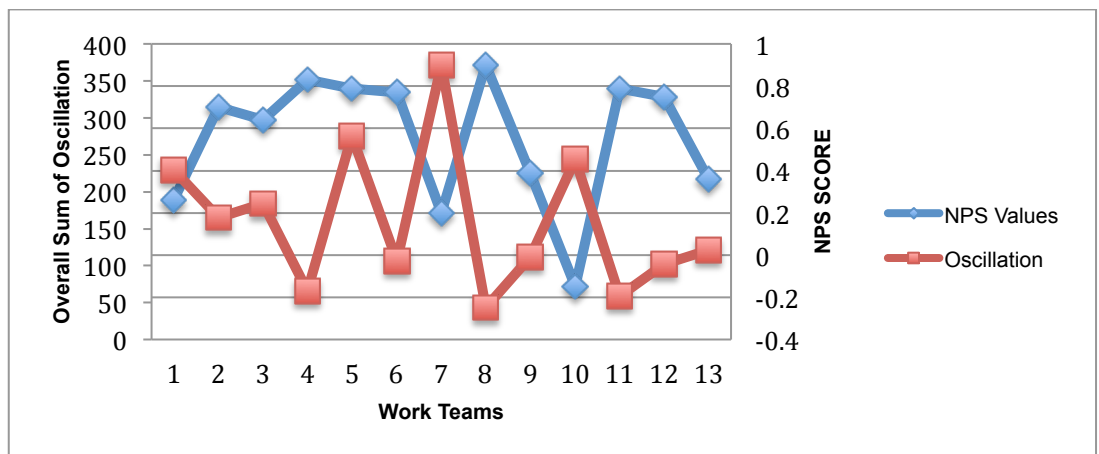

**Figure 3: Leadership Oscillation & NPS Values**

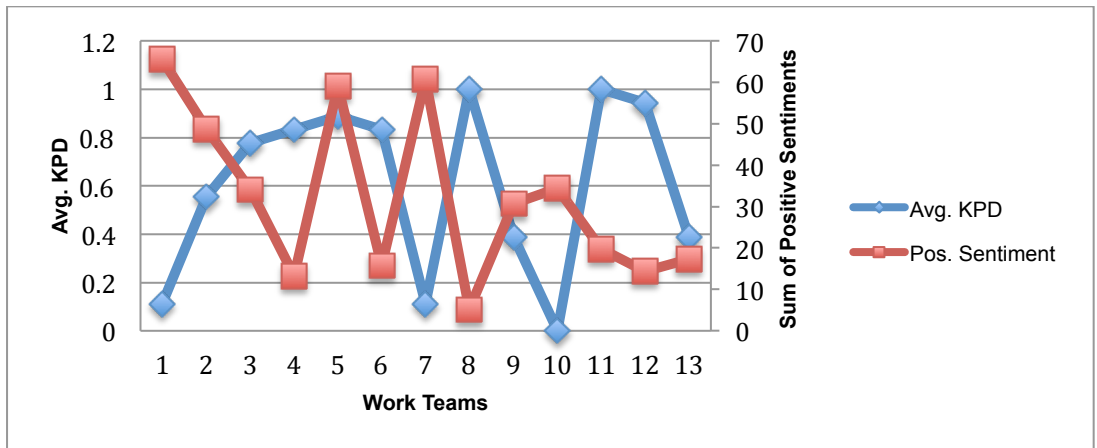
**Figure 4: Negative Correlation between Avg. KPD & Emotionality (represented as Sum of Positive Sentiments)**

Accordingly, this suggests that work teams with high variances in contribution obtain better results in client satisfaction, meaning that some team leaders are responsible for a disproportionally high amount of e-mail communication.

*Communication Behavior & Emotionality*

As was shown in prior research, work teams perform better when the communication with client employees is faster. This is tracked by calculating the median response time that is (almost) significantly correlated with KPD.

At the same time, when the e-mails' subject lines are too emotional, the client's satisfaction decreases (see Figure 4).

Therefore, hypotheses 5 and 7 are confirmed as well.

## DISCUSSION

Our empirical results capture key drivers for client satisfaction. A stable and consistent work team with clear and steady leadership that communicates fast and not over-emotional outperforms other teams.

This study thus introduces a novel Communication Score Card consisting of the eight network metrics shown in Table 1.

Compared to existing concepts in evaluating large clients' satisfaction, our approach possesses some clear advantages. By continuously tracking changes in network structure and dynamics, service providers might be able to act faster and manage projects more easily.

In addition, the analysis process can be conducted in a standardized way based on daily e-mail communication data that is already available in organizations.

In summary, on the practical level we have introduced a new communication score card (table 2) that organizations might use as a dashboard to obtain early warning signs of an impending crisis, allowing them to act proactively. On the theoretical level, we have introduced a new way to measure organizational performance with the potential to revolutionize management science by making previously immeasurable attributes measurable.

| Social Network Metric | Direction of Correlation |
|---|---|
| Group Betweenness Centrality | + |
| Group Degree Centrality | + |
| Group Density | + |
| Average new team members | − |
| Leadership Oscillation | − |
| ART (Median) | − |
| AWVCI (weighted by #actors) | + |
| Emotionality | − |

**Table 2: Network Communication Score Card**

## REFERENCES

Aral, S., & Van Alstyne, M. (2007). Network structure & information advantage. In *Proceedings of the Academy of Management Conference, Philadelphia, PA* (Vol. 3). Retrieved from http://www.chicagobooth.edu/research/workshops/orgs-markets/docs/aral-networkstructure.pdf

Bulkley, N., & Van Alstyne, M. (2006). An empirical analysis of strategies and efficiencies in social networks. *Boston U. School of Management Research Paper*, (2010-29), 4682–08.

Burt, R. S. (2001). Structural holes versus network closure as social capital. *Social capital: Theory and research*, 31–56.

Cummings, J. N., & Cross, R. (2003). Structural properties of work groups and their consequences for performance. *Social Networks*, *25*(3), 197–210.

Gloor, P. A. (2005). *Swarm creativity: Competitive advantage through collaborative innovation networks*. Oxford University Press, USA.

Gloor, P. A., Grippa, F., Putzke, J., Lassenius, C., Fuehres, H., Fischbach, K., & Schoder, D. (2012). Measuring social capital in creative teams through sociometric sensors. *International Journal of Organisational Design and Engineering*, *2*(4), 380–401.

Gloor, P. A., Laubacher, R., Dynes, S. B., & Zhao, Y. (2003). Visualization of communication patterns in collaborative innovation networks-analysis of some w3c working groups. In *Proceedings of the twelfth international conference on Information and knowledge management* (pp. 56–60).

Gloor, P. A., Paasivaara, M., Schoder, D., & Willems, P. (2008). Finding collaborative innovation networks through correlating performance with social network structure. *International Journal of Production Research*, *46*(5), 1357–1371.

Granovetter, M. (2005). The impact of social structure on economic outcomes. *The Journal of Economic Perspectives*, *19*(1), 33–50.

Jack, S. L. (2010). Approaches to studying networks: Implications and outcomes. *Journal of Business Venturing*, *25*(1), 120–137.

Kidane, Y. H., & Gloor, P. A. (2007). Correlating temporal communication patterns of the Eclipse open source community with performance and creativity. *Computational and mathematical organization theory*, *13*(1), 17–27.

Merten, F., & Gloor, P. (2010). Too Much E-Mail Decreases Job Satisfaction. *Procedia-Social and Behavioral Sciences*, *2*(4), 6457–6465.

Nemoto, K., Gloor, P., & Laubacher, R. (2011). Social capital increases efficiency of collaboration among Wikipedia editors. In *Proceedings of the 22nd ACM conference on Hypertext and hypermedia* (pp. 231–240).

Pentland, A. S. (2008). *Honest signals: how they shape our world*. MIT press.

Reichheld, F. F. (2003). The one number you need to grow. *Harvard business review*, *81*(12), 46–55.

Wasserman, S., & Faust, K. (1994). *Social network analysis: Methods and applications* (Vol. 8). Cambridge university press.